\documentclass[]{eptcs}
\usepackage{breakurl}            
\usepackage{graphicx,amsmath,amsfonts,algorithm}
\usepackage{graphics}
\usepackage{amssymb}

\def\qed{$\Box$}

\newtheorem{prop}{Proposition}\def\PRO{\begin{prop}}\def\ORP{\end{prop}}
\newtheorem{coro}{Corollary}\def\COR{\begin{coro}}\def\ROC{\end{coro}}
\newtheorem{theo}{Theorem}\def\TH{\begin{theo}}\def\HT{\end{theo}}
\def\TH{\begin{theo}}\def\HT{\end{theo}}
\newtheorem{defi}[prop]{Definition}\def\DE{\begin{defi}}\def\ED{\end{defi}}
\newtheorem{lemme}[prop]{Lemma}\def\LE{\begin{lemme}}\def\EL{\end{lemme}}
\newcommand{\AR}[2][c]{$$\begin{array}[#1]{lllllllllllllll}#2\end{array}$$}

\def\EQ#1{\begin{eqnarray}#1\end{eqnarray}}

\def\bra#1{\langle#1{|}}
\def\ket#1{| #1 \rangle}

\def\bra#1{\langle #1 |}

\title{Algebraic characterisation of one-way patterns}
\author{Vedran Dunjko 
\institute{Department of Physics, Heriot-Watt University, UK\\School of Informatics, University of Edinburgh, UK} 
\and 
Elham Kashefi 
\institute{School of Informatics, University of Edinburgh, UK}
}
\def\titlerunning{Algebraic characterisation of one-way patterns}
\def\authorrunning{Vedran Dunjko \& Elham Kashefi}

\begin{document}
\maketitle

\begin{abstract}
We give a complete structural characterisation of the map the positive branch of a one-way pattern implements. We start with the representation of the positive branch in terms of the phase map decomposition \cite{PDM}, which is then further analysed to obtain the primary structure of the matrix $\textup{M}, $ representing the phase map decomposition in the computational basis. Using this approach we obtain some preliminary results on the connection between the columns structure of a given unitary and the angles of measurements in a pattern that implements it. We believe this work is a step forward towards a full characterisation of those unitaries with an efficient one-way model implementation.
\end{abstract}

\section{Introduction}
The one-way model of quantum computation has drawn considerable attention, mainly because it suggests different physical realisations of quantum computing \cite{oneway,RBB03}. In this model quantum states are transformed using single qubit measurements on an entangled state (called \emph{open graph state}), which is prepared from an
input state by performing controlled-$Z$ operations on pairs of qubits, including
the input system and auxiliary qubits prepared in the $\ket + =\frac{1}{\sqrt{2}} \left(\ket 0 + \ket 1 \right)$ state. Quantum measurements are probabilistic in general, and can drive the computation over $2^n$ different branches, where $n$ is the number of measurements. However, there exist sufficient conditions based on the structure of the graph state where the computation can be controlled by means of single qubit corrections, dependent on the previous measurement outcomes, so that the entire computation becomes deterministic \cite{oneway,flow,gflow07,RBB03}. In such a deterministic computation, all the branches implement the same unitary map introduced by the \emph{positive branch} (also known as the post-selected branch) which corresponds to the scenario in which every measurement collapses the qubit states to pre-selected states, typically $\vert +_{\alpha_i} \rangle =\frac{1}{\sqrt{2}} \left(\ket 0 + e^{i \alpha_i} \ket 1 \right)$. 

We give a complete structural characterisation of the map the positive branch of a one-way pattern implements. The positive branch of a one-way pattern can be expressed in terms of a \textit{phase map decomposition} $R \Phi P$ \cite{PDM, QE}, which we then further analyse to obtain the primary structure of the matrix $M$ which represents $R \Phi P$ in the computational basis. The columns of $M$ can be written as:
$$
M \textup{e}_i = \varepsilon_i B_i \overrightarrow{\varphi}
$$
where $\varepsilon_i$ are complex scalars of norm one, parametrized by the measurement angles of the input qubits, $B_i$ are signs matrices, depending on the geometry of underlying open graph state, and $\overrightarrow{\varphi}$ is a vector parametrized by the measurement angles of measured auxiliary qubits.
The primary structure offers the following simple observations concerning the matrix $\textup{M}$:
\begin{itemize}
\setlength{\itemsep}{-1mm}
 \item The first column is determined only by the geometry of the open graph state and the measurement angles of the auxiliary qubits.
\item All the entries of each column are sums of complex numbers of a fixed set, possibly differing in signs.
\item The measurement angles of input qubits parametrize the global phase factors of columns of matrix M, which otherwise depend only on the geometry of the open graph state and the measurement angles of the auxiliary qubits.
\end{itemize}

Moreover we can use this characterisation to easily prove the following simple lemma about uniform determinism. Recall that a pattern is called uniform deterministic if it is deterministic for all possible angles of measurements. 
 
\LE
A pattern is uniformly deterministic if and only if it is deterministic for all possible choices of auxiliary measurement angles. 
\EL

We then proceed to meticulously dissect the $B_i$ matrices to reveal their structure given by the following decomposition:
$$
B_i= \gamma_i \Delta_i S B N \Omega_i, 
$$
where $\gamma_i$ is a sign, which depends on the adjacency of the input qubits,
$\Delta_i, S, N$ and $ \Omega_i$ are diagonal sign matrices parameterised by the adjacencies of the set of input  to the set of output qubits, the adjacency of output qubits, the adjacency of measured auxiliary qubits, and the adjacency of the set of input to the set of measured auxiliary qubits, respectively. $B$ is a full sign matrix, parametrized by the adjacency between the set of output and the set of measured auxiliary qubits.
The scalars and the matrices are given in terms of explicit functions on graphs, represented purely graph-theoretically, as adjacency matrices, and as lists of edges.
These functions have group-theoretical properties, which we feel could further be utilised to help elucidate the following open problems:
\begin{itemize}
\setlength{\itemsep}{-1mm}
\item Simulation of given unitaries directly in the one-way model, \textit{i.e.} without reference to the circuit-based model.
\item Characterisation of graph states which implement the same map in the positive branch.
\item A refined characterisation of determinism.

\item Characterisation of the pointless measurement \cite{PM} which is a key element in defining new error correcting codes.
\end{itemize}

\section{Preliminaries}
In this section we briefly review the one-way model, and present the Phase Map Decomposition \cite{PDM, QE} of one-way patterns.
A brief summary of linear algebra, and the notation used throughout this paper is given in the Appendix.  

The process of computation in the one-way model can be summarised in the following steps:
\begin{enumerate}
\item The setting up of $n$ input qubits in an input state $\vert \psi \rangle$ 
\item The addition of $m-n$ auxiliary qubits, prepared in the state $\vert + \rangle = \frac{1}{\sqrt{2}}\left( \ket 0 + \ket 1\right)$
\item The pairwise entanglement of some qubits by means of the $\wedge Z$ interaction. This interaction is represented by an open graph state, an ordered triplet $(\Gamma, I, O),$ where $\Gamma$ represents the entanglement graph (two qubits are entangled if and only if the corresponding vertices are adjacent), $I$ is the set of input qubits/vertices and $O$ is
the set of output qubits/vertices which is a subset of the auxiliary qubits. 
\item The measurement of the input qubits and non-output auxiliary qubits (which we call \textit{pure auxiliary qubits}) in the $(X, Y)$ Bloch sphere plain, that is in the basis pairs $\{ (\vert +_{\alpha_i} \rangle, \vert -_{\alpha_i} \rangle)   \},$ parametrized by a set of measurement angles $\{ \alpha_1, \ldots , \alpha_{m-n} \}$. 
Here we use the following shorthand notation: $\ket {\pm_{\alpha}} =  \frac{1}{\sqrt{2}}\left( \ket 0 \pm e^{i \alpha}\ket 1\right).$
The set $O$ corresponds to the qubits which will not be measured.
\end{enumerate}

Without loss of generality we assume input and output qubits are not overlapping. This is not a restriction, as additional auxiliary qubits can be added, which will correspond to the overlapping qubits, to which the quantum state of the overlapping qubits can be teleported. It can be easily shown these two scenarios are equivalent. As quantum measurements are generally probabilistic, the pattern implements a general completely positive map \cite{Mcal07}. The scenario in which each measurement corresponds to the projection into the state $\vert +_{\alpha_i} \rangle$ state is called the positive branch, and the positive branch realises a linear transformation of the Hilbert space of the input qubits to the Hilbert space of the output qubits. The corresponding model is called projection-based quantum computing.

We focus on the positive branch only, for this not to be a restriction, it will suffice that the graph $\Gamma$, defined by the underlying graph state, fulfils the graph-theoretical condition of having  \emph{flow} or \emph{generalised flow} \cite{flow, gflow07}, as then by means of local single qubit corrections, conditioned on sequential measurement outcomes, the entire quantum evolution of the system can be driven to be equal to the positive branch.

We will choose the labelling of qubits so that the first $n$ labels correspond to the input qubits, the following $a=m-2n$ correspond to the measured auxiliary qubits (which we will call pure auxiliaries), and the last $n$ correspond to the output qubits. We have assumed that in a given one-way pattern all the input qubits are measured first (the first round of the computation). One could easily adapt the whole discussion of this paper to the scenario where there exist no input qubits or some of the pure auxiliary qubits are also measured in the first round by labelling such qubits among the first $n$ labels. 

The measuring of a qubit in the $\{\vert \pm_\alpha \rangle \}$ basis is equivalent to first locally rotating that same qubit by the local $Z_\alpha$ unitary transformation, followed by a measurement in the $\{ \vert \pm \rangle \}$ basis. For reference reasons, here we give the matrix representations of $\wedge Z$ and $Z_\alpha$ in the computational basis:
$$
\wedge Z = \left[\begin{matrix}
1 & 0 & 0& 0 \\
0 & 1 & 0&0\\
0 & 0 & 1&0\\
0 & 0 & 0&-1
 \end{matrix}  \right] \;\;\;\;\;\; Z_\alpha = \left[\begin{matrix}
1 & 0 \\
0 & e^{i \alpha} 
 \end{matrix}  \right] 
$$
Hence, since $\wedge Z$ and $Z_\alpha$ are commuting, the projection-based computation process can be restated as follows:

\begin{enumerate}
\item The setting up of $n$ input qubits in the input state $\vert \psi \rangle$ 
\item The addition of $m-n$ auxiliary qubits, set in the state $\vert + \rangle$ 
\item The application of local $Z$ rotations to the input and  $m-2n$ auxiliary qubits, corresponding to the measurement angles  $\{ \alpha_1, \ldots , \alpha_{m-n} \}$
\item The pairwise entanglement of some qubits by means of the $\wedge Z$ interaction. This interaction is represented by an open graph state, an ordered triplet $(\Gamma, I, O),$ where $\Gamma$ represents the entanglement graph (two qubits are entangled if and only if the corresponding vertices are adjacent), $I$ is the set of input qubits/vertices and $O$ is an $n$ qubit/vertex subset of the auxiliary qubits representing the output qubits
\item The projection of the input qubits and $m-2 n$ auxiliary qubits to the $\vert + \rangle$ state
\end{enumerate}

The first and second steps above comprise an embedding of a $2^n$ dimensional Hilbert space to a $2^m$ dimensional Hilbert space which we will denote $P$ (for \textit{preparation map}), given explicitly as:
$$
P: \ket \psi \rightarrow \ket \psi \otimes \ket{+}^{\otimes (m-n)}
$$
The application of the $m-n$ local rotations implements a map which we denote $\Phi_1$:
$$
\Phi_1 = \prod_{i=1}^{m-n} \mathbf{Z}^{(i)}_{-\alpha_i}
$$
$\mathbf{Z}^{(i)}_{-\alpha_i}$ denotes an $m$-qubit unitary, which acts trivially on the composite subspaces of all qubits, except for the $i^{th}$ qubit, where it preforms the $Z_{-\alpha_i}$ rotation. Note that this is an operator on a $2^m$ dimensional Hilbert space.
We collect the entangling interactions, $\wedge Z$, into the map $\Phi_2$:
$$
\Phi_2 = \prod_{(i,j) \in \mathcal{E}} \wedge \mathbf{Z}_{i,j}
$$
where the indexing goes across the set of unordered edges $\mathcal{E}$ of the graph state given by the graph $\Gamma$: 
$$
\mathcal{E}= \left\{\left\{v_i, v_j \right\} \vert \{v_i, v_j \right\} \subseteq V(\Gamma) \}
$$
The operator $\wedge\mathbf{Z}_{i,j}$ is an $m$-qubit unitary transformation, which acts trivially on the component subspaces of all qubits, except the composite subspace of qubits $i$ and $j$, where it preforms the $\wedge Z$ transformation. We call the cumulative action of the latter two maps the \textit{Phase map} and denote it $\Phi:$
$$
\Phi = \Phi_2 \Phi_1
$$
The last step of the computation consists of projecting the first $m-n$ qubits to the state $\ket +,$ which we denote $R$ (for \textit{restriction map:})
$$
R = \bra{+}^{\otimes (m-n)} \otimes I_{2^{n}}
$$
where $I_{2^n}$ is the identity map on the $2^n$-dimensional Hilbert space.

Now, the entire process of computation in the projection-based model is represented by
$$
R \Phi P
$$
and we call this representation the \textit{Phase map decomposition} of a given unitary operator implemented in the one-way pattern \cite{PDM, QE}. Note that one can also derive directly a phase may decomposition for any unitary operators without any references to the one-way pattern \cite{PDM}.

\section{Structural characterisation of the Phase map decomposition}

Let $\{\ket i \}_{i=1}^{2^n}$ denote the standard computational orthonormal basis of a $2^n$ dimensional complex Hilbert space. Every computational basis in this representation describes a sequence of 0-1 which is the binary representation of the integer value $i-1$. Therefore $i-1$ represented in binary, encodes the choice of states $\ket 0$ or $\ket 1$ in the component single qubit state spaces. For example, the state $\ket 3$, in a four qubit setting, represents the state $\ket 0\ket 0\ket 1\ket 0$ as $(3-1)=(0010)_2$. 

Next we refine the expression $R\Phi P \ket i$ to obtain the structure of the $i^{th}$ column of the matrix which represents $R \Phi P$ in the computational basis. 

\begin{theo}\label{t-struct}
Let $R \Phi P$ be a phase map decomposition corresponding to a positive branch of a one-way pattern over $m$ qubits, with $n$ non-overlapping input and output qubits, $a=m-2n$ measured auxiliary qubits, with the set of measurement angles $\{-\alpha_1, \ldots, -\alpha_{n+a} \}$. Then, the matrix $M$ representing $R \Phi P$ is characterised with respect to columns by the following equality:

\EQ{
\label{e-struct}
M \textup{e}_i = \varepsilon_i B_i \overrightarrow{\varphi}
}
where 
\begin{itemize}
\item $ \textup{e}_i$ is the $i^{th}$ vector of the canonical basis
\item $\displaystyle{\varepsilon_i = \left( \bigotimes_{k=1}^n \left[ {1 \atop e^{i \alpha_k}} \right] , \textup{e}_{i} \right),} $ with  $(\cdot, \cdot)$ denoting the symmetric dot product
\item $\displaystyle{\overrightarrow{\varphi} =  \bigotimes_{k=n+1}^{n+a} \left[ {1 \atop e^{i \alpha_k}} \right]}$
\item $B_i$ is a matrix of signs of dimension $2^{n} \times 2^{a},$ which depends on the underlying graph state, and we call them the \textup{sign pattern matrices}
\end{itemize}
\end{theo}

{\bf Proof.} The proof is based on simple linear algebra manipulations so we put the details in the Appendix.  The main properties used are the diagonal form of both $Z_\alpha$ and $\wedge Z$
in the computational basis. The complex phases arising form the $Z_\alpha$ local rotations are collected in the $\overrightarrow{\varphi}$ vector and in the scalars $\varepsilon_i$, and the diagonal of the $\Phi_2$ entangling operation gets spread across the sign pattern matrices $B_i.$ The proof itself presents this structure of the $B_i$ matrices (see Appendix)
\EQ{\label{e-Bi}
B_i = \sum_{j=1}^{2^n}\sum_{l=1}^{2^{(m-2n)}} b_{\left[(i-1)2^{(m-n)} +(l-1)2^n + j          \right]}  \vert j \rangle   \langle l \vert 
}
These properties will be used in the following section. \qed

\vspace{0.1cm}
For simplicity, in the expression for $\overrightarrow{\varphi}$ as a numerical vector, we omit a normalising factor of $2^{-(\frac{m-2n}{2})}$, along with the scaling factor ($2^{-(m-n)}$) of the $B_i$ matrices as it has no bearing on the structure we wish to present.

A few direct consequences of Theorem \ref{t-struct} are easily checked: 
\begin{itemize}
\item The first column of M is parametrized by the measurement angles of pure auxiliary qubits only, as $\varepsilon_1 = 1$.
\item For every column $i$, the entries per row, are of the form 
\EQ{\label{e-ei}
\varepsilon_{i} (r,\overrightarrow{\varphi})
}
for some vector of signs $r$. So, entries of a column are sums of the same set of elements, varying possibly in sings only.
\item From \ref{e-ei} it is clear that every entry of every column is a sum of elements of the set of entries of the vector $\overrightarrow{\varphi}$ varying in signs, multiplied by the column's corresponding global phase factor $\varepsilon_i.$
\end{itemize} 

As mentioned before we can also prove the following simple lemma about the uniform determinism.
\LE
A pattern is uniformly deterministic if and only if it is deterministic for all possible choices of auxiliary measurement angles. 
\EL

{\bf Proof.} Due to Theorem \ref{t-struct}, the measurement angles of the input qubits appear only as global phase factors of the columns of $M$, and these global factors $\varepsilon_i$ are of norm one. Hence the choice of measurement angles of input qubits do not influence the norm of the columns.  Also regardless of the measurement angles of the input qubits (as the product of two complex numbers of norm 1 is always norm 1), the matrix $M$ is orthogonal since its columns are orthonormal. Therefore, uniform determinism can only depend on the measurement angles of the measured auxiliary qubits.\qed

The statement of Theorem \ref{t-struct} indicates a direct method for addressing problems of equalities of patterns, and of simulating a given unitary evolution of a quantum system in the one-way model. For the first problem, we have to evaluate and check the equalities of two expressions of the form of the right-hand side of \ref{e-struct}. However, that entails knowing how to construct the sign pattern matrices from given graph states. This demands further analysis of the sign pattern matrices, which will be the topic of the next two sections. 

\section{Graph-theoretical characterisation of sign pattern matrices}
In the proof of the Theorem \ref{t-struct}, the matrices $B_i$ were defined as representations of the expression 
$$
\sum_{j=1}^{2^n}\sum_{l=1}^{2^{(m-2n)}} b_{\left[(i-1)2^{(m-n)} +(l-1)2^n + j          \right]}  \vert j \rangle   \langle l \vert 
$$
in the computational basis, and in that representation $b_l$ corresponds to the the $l^{th}$ diagonal entry of the matrix representation of $\Phi_2$ in that same basis. We now link the graph theoretical aspects of the graph state defining the pattern and the above expression.

Recall that the $\wedge Z$ interaction is diagonal in the computational basis, hence the map $\Phi_2$ is diagonal in that basis as well. 
We introduce the \textit{sign parity} function $SP$ defined as $$SP(k)=(-1)^k$$ as it visually simplifies the expressions.
It was shown in \cite{PDM} that $b_l,$ the $l^{th}$ diagonal element of $\Phi_2$ is given by the following expression:
$$
b_{l} = SP\left(\sum_{(i,j)\in \mathcal{E}} x_i x_j  \right),
$$
where $x_k$ was defined as the $k^{th}$ most significant digit ($k^{th}$ digit from the left, including leading zeroes) of $l-1$ represented in binary, and $\mathcal{E}$ is an unordered list of edges of the entanglement graph state, represented by the graph $\Gamma$. It is easy to give a graph-theoretical representation for the expression for $b_l.$ It can be shown that the expression
$\sum_{(i,j)\in \mathcal{E}} x_i x_j$~, where $x_i$ and $x_j$ are functions of $l$, counts the number of edges of a vertex-induced subgraph of the graph $\Gamma,$ where the vertex set which induces the subgraph is determined by $l$. In order to explain how $l$ determines a subset of vertices, we define a function which realises this vertex determination process by an integer:
\DE
If $V$ is a set of vertices, labelled with integers $\{1, \ldots, m \}$, and $k$ is an integer in $\{1, \ldots, 2^m  \}$ then the selection function $Sel$ is defined as follows:
$Sel(V,k)$ is a subset of $V$ such that for the vertex labelled with $l$ (which we present in the subscript) $v_{l} \in V$, $v_{l} \in Sel(V,k)$ holds if and only if the $l^{th}$ most significant digit of the $m$ digit binary representation (including leading zeroes) of $k-1$ is 1.   
\ED

This function easily extends to any finite totally ordered set $O$, via an order-preserving bijection between $O$ and $\{1, \ldots, \vert O \vert \}$.
Also, we will use the expression of the form \textit{a subset of S, selected by (the integer) k} to mean precisely $Sel(S,k).$ Using the introduced terminology, $b_{l}$ is the sign parity of the number of edges of the vertex-induced subgraph of $\Gamma$ induced by a subset of the vertices of $\Gamma,$ \textit{selected by} $l$.

Now, we can direct our attention to the expression  \ref{e-Bi} and state the following proposition about the graph-theoretical characterisation of the sign pattern matrices $B_i$~.

\PRO \label{p-Bi}
Let $\Gamma$ be the a graph of a graph state, with vertices labelled by integers $\{1, \ldots, m \},$ such that the first $n$ and last $n$ correspond to input and output vertices (qubits) respectively, and the remaining $a=m-2n$ vertices correspond to the measured auxiliary (pure auxiliary) vertices (qubits) then every entry $(B_i)_{p,q}$~, is a sign parity of the number of edges of a vertex-induced subgraph of $\Gamma$, and the inducing set of vertices $V^\prime$  depends on the triplet $(i,p,q)$ as follows: 
$$
V^\prime = Sel(I, i) \cup Sel(Aux, q) \cup Sel(O, p),
$$
where $O$ denotes the subset of output vertices, $Aux$ denotes the subset of pure auxiliary vertices, and $I$ denotes the subset of input qubits. 
\ORP

{\bf Proof.} From the graph-theoretical characterisation of the diagonal elements of $\Phi_2$ the expression for $b_{l}'s$ and the binary representation of the index of $b$ in  \ref{e-Bi} it is easy to see that the Proposition holds. \qed
  
Using the terminology of the selection function we can restate this proposition in the following fashion. The $(p,q)$ entry of the sign pattern matrix $B_i$ is the sign parity of the number of edges of a vertex induced subgraph of $\Gamma$. This inducing subset is a union of subsets of the input, output and pure auxiliary vertices. The index of $B_i$ (the corresponding column of $M$) $i$ selects a subset of the input vertices. The row $p$ selects a subset of the output vertices. Finally, the column $q$ selects a subset of the pure auxiliary vertices. 

Theorem \ref{t-struct} and Proposition \ref{p-Bi} could be potentially used in address the following problems. The equality of patterns and the simulation of a given unitary. In doing so, the essential expression we need to calculate is the expression \ref{e-ei}. If we are interested in verifying the equality of two patterns, we need to calculate and compare the matrices of their phase map decompositions, given by the Theorem \ref{t-struct}. This entails calculating the dot product of the rows of the matrices $B_i$ and the vectors $\overrightarrow{\varphi}$. Similarly, if we are trying to simulate a given unitary, expressions  \ref{e-ei} which will contain variables, as we go across all entries of all columns of the matrix $M$, will form the left-hand sides of a system of equations we will have to solve (the right-hand side being the entries of the given unitary).
 
The dot product of rows of the sign pattern matrices and the vector $\overrightarrow{\varphi}$ is in general hard to evaluate, as both vectors have an exponential lengths in the number of pure auxiliary qubits. However, the vector $\overrightarrow{\varphi}$ is represented as a Kronecker product of vectors of length 2, as it corresponds to a state space vector which can be represented as a tensor product of the minimal, two-dimensional component spaces. Such a representation contains the same number of 2-dimensional vectors, as there are measured auxiliary qubits, and so is efficient. The ability to represent the rows of the sign pattern matrices in such a compact form might assist in deriving techniques for solving and evaluating such expressions efficiently.

Hence, in the following section we focus our attention to the structure of rows of the sign pattern matrices and present the decomposition theorem for the sign pattern matrices.

\section{Decomposition of the sign pattern matrices}
If we turn our attention to any row $p$ of any matrix $B_i,$ from  Proposition \ref{p-Bi} we can see that by selecting the index $i$ (equivalently, a column of $\textup M$) and a row $p$ we have  fixed a subset of the input qubits and a subset of output qubits, respectively. The $p^{th}$ row of $B_i$
is then generated by the sign parities of the numbers of edges of the vertex-induced subgraphs of $\Gamma$, where the inducing set is a union of the selected fixed sets of input and output vertices, and the entry of that row (the column of $B_i$) then selects the additional subset of the pure auxiliary vertices.

Therefore, for fixed $p$ and $i$, the corresponding row of $B_i$, which we denote by $r$, is given entry-wise by the following expression:       
\EQ{\label{e-rk}
(r)_k = SP\left( \#E\left( \Gamma_{Sel(I,i) \cup Sel(O,p) \cup Sel(Aux,k)} \right) \right)
}
where $\#E(\Gamma)$ denotes the number of edges of the graph $\Gamma,$ and for a given graph $\Gamma$ over the set of vertices $V$, and $V^\prime \subseteq V,$ $\Gamma_{V^\prime}$ denotes a vertex-induced subgraph of the graph $\Gamma$ induced by the set $V^\prime.$ In expression \ref{e-rk} only the subsets of pure auxiliary vertices change as we go across the entries of $r$.

The subgraph inducing vertex subset is expressed as a union of three subsets, two constant, and one variable.
Let us denote $A = Sel(O,p),$ $B= Sel(I, i)$ and $X = Sel(Aux, k)$. As we will be dealing with only one graph at a time, we will drop the graph designation and use the shorthand $\#E(A)$ instead of $\#E(\Gamma_A)$. Also, with $\#E(Y \leftrightarrow Z)$ we denote the number of edges joining vertices in $Y$ with vertices in $Z$ in the graph we are observing. 
It is then easy to see that 
\EQ{\label{e-E}
\#E(A \cup X\cup B) =  \#E(A) + \#E(X) + \#E(B) + \#E(B \leftrightarrow A) + \#E(A \leftrightarrow X) + \#E(B \leftrightarrow X)
}
Equality \ref{e-E} and the fact that  the sign parity function is a homomorphism from additive monoid of integers to the multiplicative monoid of integers ($SP(i+j) = SP(i)SP(j)$)
will give a basis for the decomposition of the sign parity matrices. Therefore, we can express the vector $r$ with respect to the entries as follows:
\EQ{\label{e-rk2}
(r)_k = SP\left( \#E(B)  \right) SP\left( \#E(B \leftrightarrow A) \right) SP\left( \#E(A) \right) SP\left( \#E(A \leftrightarrow X) \right) SP\left( \#E(X \right) SP\left( \#E(B \leftrightarrow X) \right)
}
Note the dependencies of the factors of the right-hand side of \ref{e-rk2} with respect to the explicit parameter $k$ of $r$, parameter $p$ which is the row selection of $B_i$ and parameter $i$ itself which is the choice of the column of $\textup{M}$ $B_i$.

\begin{enumerate}
\item $SP\left( \#E(B)  \right)$ depends on $i$ only, as it corresponds to a choice of the subset of output vertices.
\item $SP\left( \#E(B \leftrightarrow A) \right)$ depends on both $i$ and $p$, but is independent of $k$.
\item $SP\left( \#E(A) \right)$ depends on $p$ only as it corresponds to a choice of input vertices.
\item $SP\left( \#E(A \leftrightarrow X) \right)$ depends on $p$ and $k$.
\item $SP\left( \#E(X) \right)$ depends on $k$ only.
\item $SP\left( \#E(B \leftrightarrow X) \right)$ depends on $i$ and $k$. 
\end{enumerate}

We have represented the fixed row of a sign pattern matrix $r$ by its entries.
We will now represent $r$ by using vector functions, defined on graphs, as that will allow for a simple characterisation of $B_i$ matrix entries. 

First, we note that, in the list of dependencies of factors which make up an entry of $f$, the first three are constants in $k$.
The last three factors depend on $k,$ and we shall represent them as components of values (which are vectors) of two different vector functions on graphs attain.\\

We define a function on simple graphs whose set of vertices are equipped with a strict order.
\DE
Let $\Gamma$ be a simple graph, where the set of vertices is equipped with a strict order. We define $\mathcal{P}(\Gamma)$ to be a vector of signs of length $2^{\vert V\vert}$ given by the following components
$$
\left(\mathcal{P}(\Gamma)\right)_k= SP\left( \#E(Sel(V, k))  \right)
$$
for all $k= 1, \ldots, 2^{\vert V \vert}$.
\ED

Since we will often be expressing the $\mathcal{P}$ function of some vertex-induced subgraph of a graph, it is convenient to adopt a shorthand notation. If the graph $\Gamma$, which we talk about is clear, and $S$ is a subset of its set of vertices, then $\mathcal{P}(S)$ will be shorthand for $\mathcal{P}(\Gamma_S)$. Recall that $\Gamma_S$ denotes the vertex-induced subgraph of the graph $\Gamma$, induced by the set of vertices $S.$ 

The other useful function is defined on bipartite graphs.
\DE
Let $\Gamma$ be a bipartite graph with partitions $V$ and $W$, where the set of vertices is equipped with a strict order. We define $\mathcal{B}_\Gamma (V,W)$ to be a vector of signs of length $2^{\vert W \vert}$, given by the following components
$$
\left(\mathcal{B}_\Gamma(V,W)\right)_k = SP\left(\#E(V \cup Sel(W,k))  \right)
$$
for $k=1, \ldots, 2^{\vert W \vert}.$
\ED
Again, if the graph $\Gamma$ is clear from context, we will omit the subscript $\Gamma,$ and simply write $\mathcal{B}(V,W)$ instead of $\mathcal{B}_\Gamma(V,W)$.
These two functions can be explicitly defined on different representations of graphs, and these representations have potentially useful properties. We give these properties after we have given the theorem about the decomposition of the sign pattern matrices.

The row $r$ can now be expressed (as its transpose, that is as a column) using $\mathcal{B}$ and $\mathcal{P}$ functions. As the goal is to represent a general column $r$ (that is, any of the rows of any matrix $B_i$), we introduce these parameters for row $r$ - its row index $p$, and its sign pattern matrix denoted by $i$. Therefore, $r_{p,i}$ is now expressed as:
\EQ{\label{e-ri}
r_{p,i} = \gamma_i\, c^1_{p,i}\, c^2_{p}\, \cdot \left( \mathcal{B}(Sel(O,p), Aux) \odot \mathcal{P}(Aux)  \odot \mathcal{B}(Sel(I, i), Aux)\right)
}
where $I$, $Aux$ and $O$ denote the sets of input, pure auxiliary and output vertices and $\odot$ denotes the pointwise product.
The order of the components corresponds to the order of factors in the entry-wise representation of $r$ in \ref{e-rk2}.

In \ref{e-ri} $\gamma_i$ is a scalar, corresponding to the factor $SP\left( \#E(B)  \right)$ in \ref{e-rk2}. 
So we can represent it using the $\mathcal{P}$ function as $$\gamma_i= \left(\mathcal{P}(I)\right)_i$$
Also, $c^1_{p,i}$ is a constant scalar for every entry of a fixed row (hence depends on the row, and the choice of $B_i$), and corresponds to the expression $SP\left( \#E(B \leftrightarrow A) \right)$ and it can be represented using the $\mathcal{B}$ function
$$c^1_{p,i} = (\mathcal{B}(Sel(I,i),O))_p$$
Finally, $c^2_{p}$ is a constant scalar for a fixed row, and does not depend on the choice of $B_i$, and corresponds to the term $SP\left( \#E(A) \right)$. It can be represented using the function $\mathcal{P}$
$$c^2_{p} = \left(\mathcal{P}(O) \right)_p.$$

The three row-wise constant factors have been defined as components of vectors which depend on $i$ or are constant.
Then, by collecting the components across rows, and indexes $i$ we can easily note the following deconstruction of the sign pattern matrices.

\TH \label{t-struct2}
Let $V=I \cup Aux \cup O$ be the set of vertices of the graph $\Gamma$, tri-partitioned into input, auxiliary and output vertices.
Let
\begin{itemize}
\item $\gamma_i = (\mathcal{P}(\Gamma_I))_i$
\item $\Delta_i =diag(\mathcal{B}(Sel(I,i),O))$
\item $S = diag\left(\mathcal{P}(O) \right)$
\item $B = \left[  \mathcal{B}(Sel(O,1), Aux) ,\ldots ,  \mathcal{B}(Sel(O,2^{\vert O \vert}), Aux)    \right]^\tau$
\item $N=diag(\mathcal{P}(Aux))$ and
\item $\Omega_i = diag( \mathcal{B}(Sel(I, i), Aux) )$
\end{itemize}
then
$$
B_i = \gamma_i \Delta_i  S B N \Omega_i .
$$
\HT

{\bf Proof.} The origin of $\gamma_i$ is straightforward and the $\Delta_i$ and $S$ matrices are a direct consequence of the $\mathcal{P}$ and $\mathcal{B}$ function representations of the scalars $c^1_{p,i}$ and $c^2_{p}$ given above.

The $B$ matrix contains the first factor in the brackets in the expression \ref{e-ri} in each row, which is constant in $i,$ but variable in row $p.$

Matrix $N$ is the second factor in brackets in expression \ref{e-ri} spread across the diagonal of a diagonal matrix. That factor was constant in $p$ and $i$ and by presenting it as a diagonal matrix which multiplies $B$ from the right, we achieve the pointwise multiplication of each row of $B$ with that factor.
Analogous reasoning is used for the matrix $\Omega_i$ with the difference that it is variable in $i.$ \qed

Collecting the results of theorems \ref{t-struct} and \ref{t-struct2} we get the following corollary:
\COR \label{c-struct3}
Using the notation of theorems \ref{t-struct} and \ref{t-struct2} the matrix $M$ can be represented with respect to columns as:
\begin{gather}
M \textup{e}_i = \varepsilon_i  \gamma_i \Delta_i  S B N \Omega_i  \overrightarrow{\varphi} \nonumber.
\end{gather}
\ROC

While Theorem \ref{t-struct2} completely decomposes the sign pattern matrices, for an actual calculation of the presented decomposition it is useful to have the explicit forms of the functions $\mathcal{B}$ and $\mathcal{P}.$ In the following section we present different representations of these functions, and present some of their properties which could be helpful in the application of the decomposition of Corollary \ref{c-struct3}.

\section{Explicit representations of the $\mathcal{P}$ and $\mathcal{B}$ functions}
The function $\mathcal{B}$ has an elegant representation in terms of the adjacency matrix of the bipartite graph.
If $\mathcal{A}$ is the adjacency matrix of a bipartite graph, with partitions $V$ and $W,$ and all the labels of $V$ precede the labels of $W$, then it is of the block form:
$$
\mathcal{A} = \left[\begin{matrix}
0 & C \\
C^\tau & 0 
 \end{matrix}\right]
$$ 

We now define the function $\hat{\mathcal{B}} : \{0,1\}^n \rightarrow \{-1,1\}^{2^n},$
such that
\EQ{\label{e-B}
\hat{\mathcal{B}} (b_1, \ldots, b_n) = \bigotimes_{i=1}^n \left[{1 \atop SP(b_i)} \right].
}
It can be shown that if $v$ is the modulo 2 sum of the columns of $C$, then 
$$
\mathcal{B}(V,W) = \hat{\mathcal{B}}(v).
$$
That is, the $\mathcal{B}$ function can be calculated directly from the adjacency matrix of the bipartite graph in question, by using the $\hat{\mathcal{B}}$ function. Moreover, the $\hat{\mathcal{B}}$ function is a monomorphism from the group $(\{0,1\}^n, \oplus)$ to the group $(\{-1,1\}^{2^n}, \odot),$ where $\oplus$ and $\odot$ represent modulo 2 addition and pointwise multiplication, respectively.

The $\hat{\mathcal{B}}$ representation of $\mathcal{B}$ and the monomorphism property are important as described below. Given the adjacency matrix of the underlying graph we can efficiently compute a polynomial number of entries of the matrix-vector multiplication $B \overrightarrow{\varphi}$ (Corollary \ref{c-struct3}), even though the mere length of a row of $B$ is exponential in the number of auxiliary qubits. It will suffice to use the representation $\hat{\mathcal{B}}$ given on the right-hand side of \ref{e-B}, and $\overrightarrow{\varphi}$ represented in the Kronecker product form, and use the following property of the scalar product on tensor spaces:
$$
(\bigotimes_{i=1}^n X_i , \bigotimes_{i=1}^n Y_i      ) = \prod_{i=1}^n (X_i, Y_i),
$$   
when $X_i$ and $Y_i$ are of equal dimensions.

The monomorphism property also helps in the scenario where we want to calculate $B\Omega_i \overrightarrow{\varphi}$.
Since both the rows of $B$ and the diagonal of $\Omega_i$ are represented by the $\mathcal{B}$ functions, and hence by the $\hat{\mathcal{B}}$ functions, due to the monomorphism property, the pointwise product of a row in $B$ and the diagonal of $\Omega_i$ is again representable in the $\hat{\mathcal{B}}$ form, which easily reads out of the adjacency matrix, so this becomes efficiently solvable as well.

However, there remains the problem of the matrix $N,$ as what we really wish to calculate is $BN\Omega_i \overrightarrow{\varphi}$, which is represented by the $\mathcal{P}$ function. The $\mathcal{P}$ function results the sign parities of the number of edges of all subgraphs of a given graph as a binary vector.

One way to explicitly represent it is by taking the positive part of the directed adjacency matrix of the given graph $\Gamma$.
That is, we direct the graph in an arbitrary fashion, and in its directed adjacency matrix (which carries 1 and $-1$ depending on the direction of the directed edges, of the now directed graph) replace all $-1$'s with zeroes.
If $\mathcal{A}$ is that matrix then it can easily be seen that
$$
(\mathcal{P}(\Gamma))_i =SP \left( \left(\mathcal{A} \left[i\right]_2, \left[i\right]_2\right) \right),
$$
where $\left[i\right]_2$ is the binary representation of $i-1$ given as a vector.
 
If $n$ is the number of vertices, this representation takes $n^2$ binary digits on input, as they make up the $\mathcal{A}$ matrix.

An alternative representation uses $\binom{n}{2}$ binary digits in the form of an \textit{edge binary list}, which we now define. Let $E$ be an ordered set of pairs of vertices of $\Gamma$ such that the label of the first vertex in a pair is strictly smaller than the label of the second, all in all $\binom{n}{2}$ of them, and let $E$ be ordered lexicographically according to edges;
$$E = \{ (v_i, v_j) \vert v_k \in V \ \& \ i<j \}.$$
Then, for a given graph $\Gamma$, $V = V(\Gamma)$ with $\mathcal{E}$ we denote the binary vector of length $\binom{n}{2}$ such that the $i^{th}$ entry of $\mathcal{E}$ is $1$ if the $i^{th}$ pair of vertices of $E$ is adjacent in $\Gamma$ and 0 otherwise. We call this vector the \textit{edge binary list}. It is easy to see that the edge binary list uniquely characterises a simple graph.
If $\Gamma$ is a graph, and $\mathcal{E} = \left(b_1, \ldots, b_{\binom{n}{2}}\right)$ its \textit{edge binary list} then 
$\mathcal{P}\left(\Gamma\right)$ can be explicitly given as  

$$\left(\mathcal{P}\left(\Gamma\right)\right)_i = \left(\mathcal{P}\left(b_1, \ldots, b_{\binom{n}{2}}\right)\right)_i  = 
\prod_{k=1}^{\binom{n}{2}} (-1)^{b_k X_1(i,k) X_2(i,k)}, 
$$
where 
%
%$$X_1(i,k) = \left\lfloor \frac{i}{ 2^{ \left\lfloor \frac{\sqrt{8 k - 7}+1}{2}    \right\rfloor } } \right\rfloor\ \mod 2, $$ and
%$$X_2(i,k) = \left\lfloor \frac{i}{ 2^{
%\left( k -\left( { {\left\lfloor \frac{\sqrt{8 k - 7}+1}{2}  \right\rfloor} \atop  2   } \right) -1 
% \right)
%} } \right\rfloor\ \mod 2. $$
%
%
%$$X_2(i,k) = \left\lfloor \frac{i}{ 2^{
%\left( k -\binom{\left\lfloor \frac{\sqrt{8 k - 7}+1}{2}  \right\rfloor}{2}   -1 
% \right)
%} } \right\rfloor\ \mod 2. $$

$$X_1(i,k) = \left\lfloor \frac{i}{ 2^{ f(k) } } \right\rfloor\ \mod 2, $$ and

$$X_2(i,k) = \left\lfloor \frac{i}{ 2^{
\left( k -\binom{f(k)}{2}    -1 
 \right)
} } \right\rfloor\ \mod 2, $$with

$$f(k) = \left\lfloor \frac{\sqrt{8 k - 7}+1}{2}    \right\rfloor.$$

The unappealing functions $X_1$ and $X_2$ can be explained more simply. Let $(v_p, v_q)$ be the $k^{th}$ entry of the set $E$.
Then $X_1(i,k)$ is the $q^{th}$ binary digit of binary represented $i-1$, counting from the least significant digit.
With the same notation $X_2(i,k)$ is the $p^{th}$ binary digit of binary represented $i-1$, counting from the least significant digit. This representation, even though seems to be the least elegant has one significant properties. For $\mathcal{P}$ defined on edge binary lists, $$\mathcal{P} : \{0,1\}^{\binom{n}{2}} \rightarrow \{-1, 1\}^{2^n}$$ is a monomorphism from the group $\left(  \{0,1\}^{\binom{n}{2}}, \oplus \right)$ to the group $\left( \{-1, 1\}^{2^n}, \odot \right)$ where $\oplus$ denotes pointwise modulo 2 addition, and $\odot$ pointwise multiplication.

How to use this, or any other representation of the $\mathcal{P}$ function to help efficiently evaluate or express
$N \overrightarrow{\varphi},$ or $BN\Omega_i \overrightarrow{\varphi}$ in conjunction with the $\hat{\mathcal{B}}$ representation of $\mathcal{B}$ remains an open question.

\section{Discussion}
We have presented a complete structural characterisation of the positive branch of a one-way pattern in terms of its matrix representation in the computational basis. This structure was shown to be intricate and complex yet admitting a high degree of regularity. While it remains unclear how to directly use this regularity to tackle problems such as direct simulation of unitaries in one-way model or full characterisation of pointless measurements and etc., however the proposed structure clearly emphasises the importance of the entanglement. Here, entanglement plays a crucial role in the mathematical structures which arise from mathematical descriptions of the process of quantum computation; If the pure auxiliary qubits are unconnected (unentangled), the matrix $N$, of the decomposition of Theorem \ref{t-struct2}, is the identity matrix. In that case, all the entries of the matrix realised by $R \Phi P$ can be quickly evaluated, once an open graph state and the measurement angles are given. If the pure auxiliary qubits are connected, this becomes an exponential task.
We get a similar effect if we try to solve one restriction of the problem of simulating a given unitary. In this restricted problem an open graph state is given with the unitary, and it is promised that for a certain choice of angles, the positive branch will implement that unitary.
For this promise problem it can be shown that it is easily and efficiently solvable if the pure auxiliary vertices of the given graph are unconnected, for some families of graphs. Clearly, entanglement is again crucial.
It is our belief that additional work on understanding the algebraic properties of the $\mathcal{P}$ function, that is, of the graph states represented as sign patterns, may yield efficient algorithms for some instances of hard open problems in the one-way model. Such solved instances can benefit the understanding of quantum computation in general. 

\bibliographystyle{eptcs}
\bibliography{refs}

\begin{thebibliography}{1}
\providecommand{\bibitemstart}[1]{\bibitem{#1}}
\providecommand{\bibitemend}{}
\providecommand{\bibliographystart}{}
\providecommand{\bibliographyend}{}
\providecommand{\url}[1]{\texttt{#1}}
\providecommand{\urlprefix}{Available at }
\providecommand{\bibinfo}[2]{#2}
\bibliographystart

\bibitemstart{QE}
\bibinfo{author}{N.~de~Beaudrap}, \bibinfo{author}{V.~Danos},
  \bibinfo{author}{E.~Kashefi} \& \bibinfo{author}{M.~Roetteler}
  (\bibinfo{year}{2008}): \emph{\bibinfo{title}{Quadratic Form Expansions for
  Unitaries}}.
\newblock In: {\sl \bibinfo{booktitle}{Theory of Quantum Computation,
  Communication, and Cryptography Third Workshop, TQC 2008 Tokyo, Japan}},
  number \bibinfo{number}{5106} in \bibinfo{series}{Lecture Notes in Computer
  Science}.
\bibitemend

\bibitemstart{gflow07}
\bibinfo{author}{D.~Browne}, \bibinfo{author}{E.~Kashefi},
  \bibinfo{author}{M.~Mhalla} \& \bibinfo{author}{S.~Perdrix}
  (\bibinfo{year}{2007}): \emph{\bibinfo{title}{Generalized Flow and
  Determinism in Measurement-based Quantum Computation}}.
\newblock {\sl \bibinfo{journal}{New Journal of Physics}} \bibinfo{volume}{9}.
\bibitemend

\bibitemstart{flow}
\bibinfo{author}{Vincent Danos} \& \bibinfo{author}{Elham Kashefi}
  (\bibinfo{year}{2006}): \emph{\bibinfo{title}{Determinism in the one-way
  model}}.
\newblock {\sl \bibinfo{journal}{Phys. Rev. A}}
  \bibinfo{volume}{74}(\bibinfo{number}{5}), p. \bibinfo{pages}{052310}.
\bibitemend

\bibitemstart{PDM}
\bibinfo{author}{Niel deBeaudrap}, \bibinfo{author}{Vincent Danos} \&
  \bibinfo{author}{Elham Kashefi} (\bibinfo{year}{2006}):
  \emph{\bibinfo{title}{Phase map decomposition for unitaries}}.
\newblock {\sl \bibinfo{journal}{(quant-ph/0603266),}} .
\bibitemend

\bibitemstart{PM}
\bibinfo{author}{E.~Kashefi}, \bibinfo{author}{D.~Markham},
  \bibinfo{author}{M.~Mhalla} \& \bibinfo{author}{S.~Perdrix}
  (\bibinfo{year}{2009}): \emph{\bibinfo{title}{Information Flow in Secret
  Sharing Protocols}}.
\newblock {\sl \bibinfo{journal}{\emph{EPTCS}}} \bibinfo{volume}{9},
  p.~\bibinfo{pages}{87}.
\bibitemend

\bibitemstart{oneway}
\bibinfo{author}{R.~Raussendorf} \& \bibinfo{author}{H.-J. Briegel}
  (\bibinfo{year}{2001}): \emph{\bibinfo{title}{A one-way quantum computer}}.
\newblock {\sl \bibinfo{journal}{Physical Review Letters}}
  \bibinfo{volume}{86}(\bibinfo{number}{5188}).
\bibitemend

\bibitemstart{RBB03}
\bibinfo{author}{R.~Raussendorf}, \bibinfo{author}{D.{\,}E. Browne} \&
  \bibinfo{author}{H.{\,}J. Briegel} (\bibinfo{year}{2003}):
  \emph{\bibinfo{title}{Measurement-based quantum computation with cluster
  states}}.
\newblock {\sl \bibinfo{journal}{Physical Review A}} \bibinfo{volume}{68}, p.
  \bibinfo{pages}{022312 [32 pages]}.
\bibitemend

\bibitemstart{Mcal07}
\bibinfo{author}{V.Danos}, \bibinfo{author}{E.Kashefi} \&
  \bibinfo{author}{P.Panangaden} (\bibinfo{year}{2007}):
  \emph{\bibinfo{title}{The Measurement Calculus}}.
\newblock {\sl \bibinfo{journal}{Journal of ACM}} \bibinfo{volume}{54}, p.
  \bibinfo{pages}{8 [45 pages]}.
\bibitemend

\bibliographyend
\end{thebibliography}

\section{Appendix}
\subsection{Summary of notation}
Here we present a brief summary of the notation used throughout the paper. The algebra used is presented in the Dirac notation.
\paragraph{Qubit states}
A \textit{qubit} is represented by a two-dimensional complex Hilbert space, called the qubit's state space. A \textit{qubit state} is a vector of unit length in the qubit's state space. The \textit{state space of an ensemble of qubits} is represented by the tensor product of the component state spaces, and the \textit{state of an ensemble of qubits} is a vector of unit length in the state space of the ensemble. With $\ket 0$ and $\ket 1$ we denote unit orthonormal vectors in the state space of a qubit, and they constitute the \textit{standard computational basis} of a qubit. $\ket{\pm_\alpha}$ denotes a vectors parameterised by the real angle $\alpha$ (and the choice of $+$ or $-$) defined with respect to the computational basis vectors as $$\ket{\pm_\alpha} = \dfrac{1}{\sqrt{2}}\left(\ket 0 \pm e^{i \alpha} \ket{1}\right).$$ When $\alpha = 0$, we simply write $\ket \pm.$
\paragraph{Unitary transformations} $Z_\alpha$ denotes a family of \textit{phase shift unitary transformations}, parametrized by the real angle $\alpha$, represented in the computational basis with the following matrix:
$$
 Z_\alpha = \left[\begin{matrix}
1 & 0 \\
0 & e^{i \alpha} 
 \end{matrix}  \right] .
$$
When the $Z_\alpha$ rotation is applied to the $i^{th}$ qubit of an ensemble of $m$ qubits, the transformation of the state space of the ensemble is denoted with $\mathbf{Z}^{(i)}_{\alpha},$ which can be given explicitly with
$$
\mathbf{Z}^{(i)}_{\alpha} = I^{\otimes (i-1)} \otimes Z_{\alpha} \otimes I^{\otimes (m-i).} 
$$
Here, $I$ denotes the identity operator on a single qubit state space. $\wedge Z$ denotes a unitary transformation on the state space of two qubits. In the computational basis it is given by the following matrix:

$$
\wedge Z = \left[\begin{matrix}
1 & 0 & 0& 0 \\
0 & 1 & 0&0\\
0 & 0 & 1&0\\
0 & 0 & 0&-1
 \end{matrix}  \right].$$
Note that this operator cannot be represented as a tensor product of single qubit transformations. Hence, it can be used to create \textit{entangled states,} which are multi-qubit states which cannot be represented as tensor products of single-qubit states.
 
When the $\wedge Z$ transformation is applied to the component state subspace of the $i^{th}$ and $j^{th}$ qubit of an ensemble of $m$ qubits, the transformation of the entire ensemble is denoted with $\wedge \mathbf{Z}_{i,j}.$
The eigenvectors of the $\wedge \mathbf{Z}_{i,j}$ transformation are the vectors of the computational basis of the ensemble, with eigenvalue $-1$ if both the $i^{th}$ and $j^{th}$ qubit are in the state $\ket 1$ and eigenvalue $1$ otherwise.
\paragraph{Miscellaneous}
\begin{itemize}
\item $e$ denotes the basis of the natural logarithm.
\item $\textup{e}_i$ denotes the $i^{th}$ vector of the canonical basis, i.e. a vector with entries $0$ everywhere, except a $1$ at the $i^{th}$ entry.
\item $\otimes$ represents the tensor product. $X^{\otimes n}$ denotes the $n$-th tensor power of $X,$ explicitly 
$$
X^{\otimes n} = \overbrace{X \otimes \cdots \otimes X}^{\mbox{n\ times}}.
$$
The tensor product of matrices (and also numerical vectors, as they are isomorphic to single row or column matrices) is called the Kroenecker and defined explicitly as follows:\\

If $A$ is an $m$-by-$n$ matrix and $B$ is a $p$-by-$q$ matrix, then the Kronecker product $A\otimes B$ is the block matrix 
$$
A\otimes B = \begin{bmatrix} a_{11} B &  \cdots & a_{1n}B \\ \vdots &  \ddots &  \vdots \\ a_{m1} B &  \cdots &  a_{mn} B \end{bmatrix}.$$
\item $SP(n),$ for an integer $n$ denotes the \textit{sign parity} function defined as
$$
SP(n)=(-1)^n. 
$$
\item $X^\tau$ denotes the transpose of the matrix (or vector) $X.$
\item $(\cdot, \cdot)$ denotes the symmetric dot product;
If $X=\left[x_1, \ldots, x_n \right]^\tau$ and $Y=\left[y_1, \ldots, y_n \right]^\tau$ are vectors, then 
$$
(X,Y) = \sum_{i=1}^{n} x_i y_i.
$$
\item If $\Gamma$ is a graph, and $A\subseteq V(\Gamma)$ a subset of the vertices of $\Gamma$, $\Gamma_A$ denotes the vertex-induced subgraph of the graph $\Gamma$ induced by the set of vertices $A.$
\item If $\Gamma$ is a graph $\#E(\Gamma)$ is the number of its vertices, i.e. $\#E(\Gamma) = \vert E(\Gamma) \vert$. If $\Gamma_A$ is a subgraph of $\Gamma,$ the graph designation can be dropped and $\#E(A)$ denotes $\#E(\Gamma_A).$ If $A$ and $B$ are disjoint subsets of the vertices of the graph $\Gamma$ $\#E(A \leftrightarrow B)$ denotes the number of edges connecting the vertices in $A$ to vertices in $B$ in the graph $\Gamma.$
\item $\oplus$ denotes the modulo 2 addition. If $X=\left[x_1, \ldots, x_n \right]^\tau$ and $Y=\left[y_1, \ldots, y_n \right]^\tau$ are vectors of integers, then 
$$
X \oplus Y = \left[x_1 \oplus y_1, \ldots, x_n \oplus y_n  \right]^\tau.
$$
\item $\odot$ denotes the pointwise vector product; If $X=\left[x_1, \ldots, x_n \right]^\tau$ and $Y=\left[y_1, \ldots, y_n \right]^\tau$ are vectors, then 
$$
X \odot Y = \left[x_1 \odot y_1, \ldots, x_n \odot y_n  \right]^\tau.
$$
\end{itemize}

\subsection{Proof of Theorem \ref{t-struct}}
Let $R \Phi P$ be the phase map decomposition \cite{PDM} of the positive branch of a one-way pattern over $m$ qubits, $n$ of which are input, $n$ output, and $a=m-2n$ are pure auxiliary qubits.
Also let $\ket i$ be a vector of the standard computational basis. Then we can directly derive the following: 

\AR{
R \Phi P \vert i \rangle &=& I_{2^n} R \Phi P \vert i \rangle \\\\
&=&\sum_{j=1}^{2^n} \vert j \rangle \langle j \vert  R \Phi P \vert i \rangle \\\\
&=& \sum_{j=1}^{2^n}\vert j \rangle \langle j \vert R \Phi \left( \vert i \rangle \otimes \vert + \rangle^{\otimes m-n} \right) \\\\
&=& \sum_{j=1}^{2^n} \vert j \rangle  \langle j \vert R \Phi_2 \Phi_1 \left( \vert i \rangle \otimes \vert + \rangle^{\otimes m-n} \right) \\\\
&=& \left( \left(\otimes_{k=1}^{n}\langle +_{\alpha_k} \vert \right) \vert i \rangle\right) \sum_{j=1}^{2^n} \vert j \rangle \langle j \vert R \Phi_2 \left( \vert i \rangle \otimes \left(\otimes_{k=n+1}^{m-n }\vert +_{\alpha_k} \rangle \right) \otimes \vert + \rangle^{\otimes n}\right)
}
For clarity reasons we temporarily omit the row-constant scalar $\left( \left(\otimes_{k=1}^{n}\langle +_{\alpha_k} \vert \right) \vert i \rangle\right)$

\AR{
&=& \sum_{j=1}^{2^n} \vert j \rangle \langle j \vert R \Phi_2 \left( \vert i \rangle \otimes \left(\otimes_{k=n+1}^{m-n }\vert +_{\alpha_k} \rangle \right) \otimes \vert + \rangle^{\otimes n}\right)\\\\
&=&\sum_{j=1}^{2^n} \vert j \rangle \langle j \vert \langle + \vert^{\otimes (m-n)}\otimes I_{2^n} \Phi_2 \left( \vert i \rangle \otimes \left(\otimes_{k=n+1}^{m-n }\vert +_{\alpha_k} \rangle \right) \otimes \vert + \rangle^{\otimes n}\right)\\\\
&=&\sum_{j=1}^{2^n} \vert j \rangle \left( \langle + \vert^{\otimes (m-n)}\otimes  \langle j \vert \right)  \Phi_2 \left( \vert i \rangle \otimes \left(\otimes_{k=n+1}^{m-n }\vert +_{\alpha_k} \rangle \right) \otimes \vert + \rangle^{\otimes n}\right)
}
We note that $\Phi_2= \sum_{l=1}^{2^m} b_l \vert l \rangle \langle l \vert$ where $b_l$ is the $l^{th}$ diagonal element of the diagonal matrix $\Phi_2$,

\EQ{
\nonumber &=&\sum_{j=1}^{2^n} \vert j \rangle \left( \left( \langle + \vert^{\otimes (m-n)}\otimes  \langle j \vert \right)  \sum_{l=1}^{2^m} b_l \vert l \rangle \langle l \vert \right) \left( \vert i \rangle \otimes \left(\otimes_{k=n+1}^{m-n }\vert +_{\alpha_k} \rangle \right) \otimes \vert + \rangle^{\otimes n}\right)
\\&=& \sum_{j=1}^{2^n} \vert j \rangle  \left( \sum_{l=1}^{2^{(m-n)}} b_{\left[ (l-1)2^n + j          \right]}   \langle l \vert\langle j \vert \right) \left( \vert i \rangle \otimes \left(\otimes_{k=n+1}^{m-n }\vert +_{\alpha_k} \rangle \right) \otimes \vert + \rangle^{\otimes n}\right)
\\ &=& \nonumber \sum_{j=1}^{2^n} \vert j \rangle  \left( \sum_{l=1}^{2^{(m-n)}} b_{\left[ (l-1)2^n + j          \right]}   \langle l \vert\langle j \vert  \left( \vert i \rangle \otimes \left(\otimes_{k=n+1}^{m-n }\vert +_{\alpha_k} \rangle \right) \otimes \vert + \rangle^{\otimes n}\right)\right) 
\\&=& \sum_{j=1}^{2^n} \vert j \rangle  \left( \sum_{l=1}^{2^{(m-n)}} b_{\left[ (l-1)2^n + j          \right]}   \langle l \vert  \left( \vert i \rangle \otimes \left(\otimes_{k=n+1}^{m-n }\vert +_{\alpha_k} \rangle \right)\right)\right) 
\\&=& \nonumber \sum_{j=1}^{2^n} \vert j \rangle  \left( \sum_{l=1}^{2^{(m-2n)}} b_{\left[(i-1)2^{(m-n)} +(l-1)2^n + j          \right]}   \langle l \vert   \left(\otimes_{k=n+1}^{m-n }\vert +_{\alpha_k} \rangle \right)\right)
\\&=&\nonumber \sum_{j=1}^{2^n}\sum_{l=1}^{2^{(m-2n)}} b_{\left[(i-1)2^{(m-n)} +(l-1)2^n + j          \right]}  \vert j \rangle   \langle l \vert   \left(\displaystyle\otimes_{k=n+1}^{m-n }\vert +_{\alpha_k} \rangle \right) 
}
So now we summarise the entire expression (reintroducing the omitted scalar):
$$
R \Phi P \vert i \rangle=  \left( \left(\bigotimes_{k=1}^{n}\langle+_{\alpha_k} \vert \right) \vert i \rangle\right)\sum_{j=1}^{2^n}\sum_{l=1}^{2^{(m-2n)}} b_{\left[(i-1)2^{(m-n)} +(l-1)2^n + j          \right]}  \vert j \rangle   \langle l \vert   \left(\bigotimes_{k=n+1}^{m-n }\vert +_{\alpha_k} \rangle \right).
$$

For simplicity, in $(8)$ we omitted a global scaling factor of $2^{-( \frac{m-n}{2})}$, brought about by the scalar products $\bra{+}^{\otimes(m-n)} \bra{j} \ket{l}$ where they are non-zero, and in $(9)$ the global scaling factor $2^{-( \frac{n}{2})}$, caused by the product $\langle j \ket{+}^{\otimes n}$. The overall (omitted) scaling factor is $2^{-\frac{m}{2}}$.  

The expression $ \left( \left(\bigotimes_{k=1}^{n}\langle+_{\alpha_k} \vert \right) \vert i \rangle\right)$ is a scalar which depends on the column $i$, and we denote it by~$\varepsilon_i$, also let
\AR{
B_i = \sum_{j=1}^{2^n}\sum_{l=1}^{2^{(m-2n)}} b_{\left[(i-1)2^{(m-n)} +(l-1)2^n + j          \right]}  \vert j \rangle   \langle l \vert 
}
be a $2^n \times 2^{(m-2n)}$ matrix expressed in the computational basis that depends on the choice of column $i$ with entries in $\{-1,1\}$. Finally, denote the numeric representation in the computational basis of the vector $\left(\bigotimes_{k=n+1}^{m-n }\vert +_{\alpha_k} \rangle \right)$ with $\overrightarrow{\varphi}$, which is independ of the choice of the column. It corresponds to the quantum state of the auxiliary qubits after the local $Z_\alpha$ rotations, but before the entanglement procedure.
The entire expression can then be rewritten in matrix notation as:
$$
M\textup{e}_i = \varepsilon_i B_i \overrightarrow{\varphi},
$$
where $\textup{e}_i$ is the $i^{th}$ vector of the canonical basis.\qed 

\end{document}